\documentclass[aps,prc,twocolumn,superscriptaddress,floatfix,nofootinbib,showpacs]{revtex4}

\usepackage{mathrsfs}
\usepackage{amsmath}
\usepackage{enumerate}
\usepackage{graphicx}
\usepackage{subfigure}
\usepackage{float}
\usepackage{slashed}
\usepackage{color}
\usepackage{CJK}

\begin{document}


\begin{CJK*}{GB}{}

\parindent=24pt

\title{Temperature Dependence of the Effective Bag Constant and
the Radius of a Nucleon in the Global Color Symmetry Model of QCD }

\author{Yuan Mo}
\affiliation{Department of Physics and State Key Laboratory of
Nuclear Physics and Technology, Peking University, Beijing 100871,
China}
\author{Si-xue Qin}
\affiliation{Department of Physics and State Key Laboratory of
Nuclear Physics and Technology, Peking University, Beijing 100871,
China}
\author{Yu-xin Liu}
\email[Corresponding author: ]{yxliu@pku.edu.cn}
\affiliation{Department of Physics and State Key Laboratory of
Nuclear Physics and Technology, Peking University, Beijing 100871,
China} \affiliation{Center of Theoretical Nuclear Physics, National
Laboratory of Heavy Ion Accelerator, Lanzhou 730000, China}

\begin{abstract}
We study the temperature dependence of the effective bag constant,
the mass, and the radius of a nucleon in the formalism of the simple
global color symmetry model in the Dyson-Schwinger equation approach
of QCD with a Gaussian-type effective gluon propagator. We obtain
that, as the temperature is lower than a critical value, the
effective bag constant and the mass decrease and the radius
increases with the temperature increasing. As the critical
temperature is reached, the effective bag constant and the mass
vanish and the radius tends to infinity. At the same time, the
chiral quark condensate disappears. These phenomena indicate that
the deconfinement and the chiral symmetry restoration phase
transitions can take place at high temperature. The dependence of
the critical temperature on the interaction strength parameter in
the effective gluon propagator of the approach is given.
\end{abstract}


\maketitle

\end{CJK*}

\section{Introduction}

The phase transitions of quantum chromodynamics (QCD), for example
the evolution between chiral symmetry breaking and its restoration,
the color (or simply quark ) confinement and deconfinement, have
been the most active topics in nuclear and particle physics in
recent years~\cite{LongRangePlan}.  Even though recent investigation
has provided hints that the phase transitions can be driven by the
intrinsic characteristics, such as the running coupling strength and
the current quark mass, of the system (see, for example,
Refs.~\cite{Craig1,Reinhard,YuanWei,ChangLei,Pennington2}),
the more promising and much better believed is that the QCD may
undergo phase transitions into a chirally symmetric and color
deconfined phase at high temperature and/or
density~\cite{deconfinement}.

To demonstrate the phase transitions, one usually implements the
variation behaviors of not only the features of QCD vacuum and the
strong interaction matter but also the properties of hadrons at
finite temperature and/or density. On theoretical side, one needs in
principle QCD, which has been widely accepted as the fundamental
theory of strong interaction, to carry out the investigation.
However, as a basic theory, QCD still suffers from difficulties in
the low energy region, which relates directly to strong interaction
matter and hadrons.
Then, besides the approaches of lattice simulations, QCD sum rules,
instanton model(s) and Dyson-Schwinger equations and several models,
such as the bag model~\cite{bag-models}, quark-meson coupling model
(QMC)~\cite{QMC1,QMC2,QMC3}, Nambu-Jona-Lasinio (NJL)
model~\cite{NJL}, Polyakov-loop improved NJL model~\cite{PNJL},
global color symmetry model (GCM)~\cite{GCM1}, and so on, have been
developed.
The NJL model has been widely used since it preserves the feature of
chiral symmetry and its dynamical breaking and is easy to carry out
numerical calculation. Even though, with the Polyakov-loop
improvement, the quark confinement effect is included at statistical
level, the commonly accepted one still only takes into account the
point (contact) interactions among quarks. The bag model is the one
which handles hadrons as bubbles of perturbative vacuum immersed in
the physical vacuum. However, all nonperturbative physics is
included in a quantity---bag constant, which is dealt with a
phenomenological parameter in the model. And the QMC model involves
the similar problem.
For the GCM, since it can take the result of the Dyson-Schwinger
equation (DSE)~\cite{DSE-Roberts} approach as input, it manifests
well the properties of chiral symmetry and its dynamical breaking.
Because the bag constant in the model is taken as the difference
between the energy densities of the perturbative and the physical
vacuums, the color confinement effect is also handled well in some
sense. The GCM is then believed to be a quite sophisticated model
which involves as many characteristics of QCD as possible. And the
NJL model, the QMC model, and the bag model can be treated as the
special cases of the GCM.

Due to its solid QCD foundation, the GCM has been widely taken to
study not only the properties of nucleon and some
mesons~\cite{GCM1,GCM2,GCMsoliton1,GCMsoliton2} but also the QCD
vacuum structure~\cite{ZHS200123,Zhangzhao1}. It has also been
extended to investigate the properties of strong interaction matter
at finite temperature and/or density and those of some hadrons in
the
matter~\cite{Johnson1997,Blaschke,LYX200135,LYX2003,LYX2005,Zhangzhao2}.
For the property of nucleon in finite density matter, it has been
discussed with various models for the effective gluon propagator in
the DSE and the variation behaviors of the mass, the radius and the
bag constant of the nucleon have been given
explicitly~\cite{LYX200135}. However, in the case of finite
temperature, only the changing feature of the bag constant has been
discussed with the Munczek-Nemirovsky model~\cite{MNmodel} for the
effective gluon propagator of the DSE~\cite{Blaschke}. We will then,
in this paper, discuss some of the properties of a nucleon at finite
temperature with a sophisticated effective gluon propagator in the
DSE.

The paper is organized as follows. In Sec. II, we describe briefly
the formalism of the GCM soliton model. In Sec. III, we describe the
algorithm to carry out the numerical calculation of the GCM soliton
at finite temperature and the obtained results. Finally we give a
brief summary and some remarks in Sec. IV.

\section{Brief Description of GCM}

The original action in the global color symmetry model (GCM) defined
in Euclidean metric is expressed as~\cite{GCM1}
\begin{eqnarray}
S_{GCM}(\bar{q},q) & \!\! = \!\! \! & \int \! {d^4x
\bar{q}(x)(i\gamma\cdot p + m_{0}^{} )q(x)}   \nonumber \\  & &
+\frac{g^2}{2}\!\! \iint \!\! {d^4xd^4yj^{a}_{\mu}(x)
D^{ab}_{\mu\nu}(x-y) j^{b}_{\nu}(y)}\, , \;\;
\end{eqnarray}
where $j^a_\mu(x)=\bar{q}(x)\frac{\lambda^\mu}{2}\gamma_{\mu}q(x)$
is the local quark current, $D^{ab}_{\mu\nu}(x-y)$ is the full gluon
propagator, $m_{0}^{}$ is the current quark mass, $g$ is the
quark-gluon coupling constant. The Euclidean metric is such that $a
\cdot b=a_{\mu} b_{\mu}$, and
$\{\gamma_{\mu},\gamma_{\nu}\}=2\delta_{\mu\nu}$.
Taking the gluon propagator to be color diagonal in the Feynman-like
gauge, i.e., $D^{ab}{\mu \nu} (x-y) = \delta^{ab} \delta_{\mu \nu}
D(x-y)$, and applying the Fierz transformation to reorder the quark
fields, one can rewrite the action as
\begin{eqnarray}
S_{GCM}[B^\theta(x,y))] & \! = \!\! & \!
\iint{d^{4}xd^{4}y}\bar{q}(x)[(i \gamma \cdot p + m_{0}^{} )
\delta(x-y)
 \nonumber   \\   &  &
 \qquad + \Lambda^{\theta} B^\theta(x,y)]q(x)  \nonumber   \\  & &
 +\iint{d^{4}x d^{4}y\frac{B^{\theta}(x,y)B^{\theta}(y,x)}{2g^{2} D(x-y)} }  \, ,
\end{eqnarray}
where $\{\Lambda^{\theta} \} $ are direct products of Lorentz,
flavor and color matrices of quarks which produce the scalar,
vector, and pseudoscalar terms labeled by $\theta$.
$B^{\theta}(x,y)$ are bilocal Bose fields.
Theoretically, it can be proved that the GCM is valid in any gauge
even though one takes the Feynman-like gauge in deriving the above
expression~\cite{GaugeInd}.

By integrating the quark fields, one gets the action
\begin{eqnarray}
S_{GCM}[B^\theta(x,y))]& \! = \! & -Tr \ln G^{-1}[B^\theta(x,y)]
\nonumber
\\  & &
+\! \iint \!\!
{d^4xd^4y\frac{B^\theta(x,y)B^\theta(y,x)}{2g^2D(x-y)}} \, , \;\;\;
\end{eqnarray}
where the inverse of the quark propagator can be written as
\begin{eqnarray}
G^{-1}(x,y)=( i \gamma \cdot p + m_{0}^{} )\delta(x-y) +
{\Lambda^{\theta}} B^{\theta}(x,y) \, .
\end{eqnarray}

Generally, the bilocal fields can be expanded as
\begin{eqnarray}
B^{\theta}(x,y)=B_{0}^{\theta}(x,y) +
\sum_{i}\Gamma_{0}^{\theta}(x,y)\phi_{i}^{\theta}
\left(\frac{x+y}{2}\right) \, ,
\end{eqnarray}
where the first term is the translation invariant vacuum
configuration. The second term stands for the fluctuations of the
vacuum which can be identified as effective meson fields since the
$\theta$ stands for the quantum number of Bose fields. In the lowest
order, one takes the Goldstone mode, $\phi_{0}^{\theta} =
\{\sigma,\vec\pi\}$, which is thought of as the most important low
energy degree of freedom. The vacuum configuration can be determined
by the saddle point condition $\delta S_{GCM}/\delta B_{0}^{\theta}
= 0$. One has then
\begin{eqnarray}
B_{0}^{\theta}(x,y)=g^{2}D(x-y) tr[ G(y,x){\Lambda^{\theta}} ] \, ,
\end{eqnarray}
and the quark self-energy $\Sigma(x-y) = {\Lambda^{\theta}}
B_{0}^{\theta}(x,y)$. The equation of quark self-energy in momentum
space coincides with that of the truncated Dyson-Schwinger equation
(DSE)
\begin{widetext}
\begin{eqnarray}
\Sigma(p)=\int{d^4x{\Lambda^{\theta}} B_{0}^{\theta}(x,y) e^{iq\cdot x}} 
=g^2\int{\frac{d^4q}{(2\pi)^4}t_{\mu\nu}D(p-q)\frac{\lambda^a}{2}
\gamma_{\mu}\frac{1}{i\gamma\cdot(q+m)+\Sigma(q)}\gamma_{\mu}
\frac{\lambda^a}{2}} \, ,
\end{eqnarray}
\end{widetext}
where $t_{\mu\nu}=\delta_{\mu\nu}-k_{\mu}k_\nu/k^2$, with $k=p-q$,
$\gamma^\mu$ is the color SU(3) matrix. Generally, the quark
self-energy function can be decomposed as
\begin{eqnarray}
\begin{split}
\Sigma(p)&=S^{-1}(p)-S^{-1}_0(p) \, , \\
S^{-1}(p)&=i\gamma\cdot p\ A(p^2)+B(p^2) \, ,\\
S^{-1}_0(p)&=i\gamma\cdot p + m_{0}  \, ,
\end{split}
\end{eqnarray}
and $A(p^2)$, $B(p^2)$ are scalar functions of $p^2$.

Recalling the configuration of the bilocal fields in Eq.~(5) and
considering the Bethe-Salpeter amplitude of the mesons and the
partial conservation of axial-vector current, one can
prove~\cite{GCM3} that, when considering the most important low
energy degree of freedom, i.e., the Goldstone mode
$\phi_{0}^{\theta} = \{\sigma,\vec\pi\}$,  Eq.~(5) can be rewritten
as
\begin{widetext}
\begin{eqnarray}
{\Lambda^{\theta}} [B^\theta(x,y)-B_0^\theta(x,y)] =
B(x-y)\left[\sigma(\frac{x+y}{2})
+i\gamma_5\vec{\tau}\cdot\vec{\pi}(\frac{x+y}{2})\right] \, ,
\end{eqnarray}
\end{widetext}
where $B$ is just the scalar part of the inverse of the quark
propagator which can be determined by solving the quark's DSE.

With a nontopological-soliton ans\"{a}tz\cite{FL778}, the action of
the GCM soliton with quarks in chiral limit ($m_{0}^{} = 0$) can be
given~\cite{GCM1,GCMsoliton1,GCMsoliton2} as
\begin{eqnarray}
S_{GCM}\! & \!= & \!\! \overline{q} \{ i \gamma \cdot p - \alpha
[\sigma(x)+i\vec{\pi}(x) \cdot \vec{\tau}\gamma_5 ] \} q \nonumber \\
& & \!\!\! + \! \int[\frac{f_{\sigma}^2}{2} (\partial_{\mu}\sigma)^2
+ \frac{f_{\pi}^2}{2} (\partial_{\mu} \vec{\pi})^2 \! - \!
V(\sigma,\pi)]d^{4}z  \, , \;\;
\end{eqnarray}
 with
\begin{eqnarray}
V(\sigma,\pi) \! & \! \approx \! & \! - 12 \int \frac{d^{4}
p}{(2\pi)^4} \left\{ \ln \left[\frac{A^{2}(p) {p}^{2} +
(\sigma^2+\vec{\pi}^2)B^{2}(p)}{A^{2}(p) p^{2} + B^{2}(p) } \right]
\right. \nonumber \\      
& & - \left. \frac{(\sigma^2+\vec{\pi}^2-1) B^{2}(p)}{A^{2}(p) p^{2}
+ B^{2}(p) } \right\}\, ,
\end{eqnarray}
and the quark meson coupling constant $\alpha$ is given as
$$ \alpha(x) = \int \frac{d^{4} p}{(2 \pi)^4} B(p) e^{-i
p \cdot x} \, .$$
 It is evident that such a quark meson coupling constant is just
the vacuum configuration of the bilocal fields and is independent of
the meson fields.

With the stationary condition of the soliton, one has the equations
of motion for the quarks and mesons as
\begin{equation}
\left\{i \gamma \cdot p - \alpha [ \sigma(x) + i \vec{\pi}(x) \cdot
\vec{\tau}\gamma_5 ] \right\} q = 0 \, ,
\end{equation}
\begin{equation}
\label{sigmaEq} -\vec\nabla^2 \sigma(\vec{r})+\frac{\delta V}{\delta
\sigma(\vec{r})}+ Q_\sigma (\vec{r})  = 0 \, ,
\end{equation}
\begin{equation}  \label{piEq}
-\vec\nabla^2 \vec{\pi}(\vec{r}) + \frac{\delta V}{\delta \vec\pi
 (\vec{r})} + Q_{\vec\pi} (\vec{r})  =  0 \, ,
\end{equation}
where $Q_\sigma$ and $Q_{\vec\pi}$ are the source terms contributed
from the valence quarks, and can be written as
\begin{eqnarray}
\label{quarksource_sigma} Q_\sigma(\vec{R}) & = &
\sum\limits_{j=1}^3 \frac{1}{ Z_j} \int d^3x d^3y
\bar{u}_j(\vec{x}) B(\vec{x}-\vec{y})   \nonumber \\
& & \times \delta(\frac{\vec{x} + \vec{y}}{2}-\vec{R}) u_j(\vec{y}) \, ,   \\
 \label{quarksource_pi}
Q_{\vec\pi}(\vec{R}) & = & \sum\limits_{j=1}^3 \frac{1}{Z_j} \int
d^3x d^3y \bar{u}_j(\vec{x}) B(\vec{x}-\vec{y}) i\gamma_5
\vec{\tau} \nonumber \\
& & \delta(\frac{\vec{x}+\vec{y}}{2}-\vec{R}) u_j(\vec{y}) \, ,
\end{eqnarray}
with $Z_{j}$ being the renormalization constant~\cite{GCMsoliton1}
\begin{equation}
Z_j=-\int d^3p d^3q \bar{u}_j(\vec{p}) \frac{\partial
G^{-1}(i\epsilon_j;\vec{p},\vec{q})}{\partial \epsilon_j}
u_j(\vec{q})  \,  .
\end{equation}

The quark field and $\sigma$, $\pi $ meson fields can be determined
by solving Eqs.~(12)-(14) self-consistently. As a consequence, the
corresponding eigenenergies can be obtained. It is apparent that the
meson fields corresponding to the vacuum configuration can be simply
taken as $\sigma =1$, $\pi = 0$ due to the (normalized with
$f_{\pi}^{2}$) restriction $\pi ^2 + \sigma ^2 =1$. The vacuum
configuration is a minimum of $V(\sigma,\pi)$ and $V(1,0)=0$.
In light of the nontopological-soliton ans\"{a}tz\cite{FL778,GCM1},
one can approximate the soliton as a chiral bag with bag constant
\begin{widetext}
\begin{equation}
{\mathscr{B}}  =  V(\sigma,\vec{\pi}) - V(1,0)  
  =  12 \int
\!\!\frac{d^4 p}{(2\pi)^4} \left\{\ln \left[ \frac{A^{2}(p)p^{2} \!
 + \! B^{2}(p)}
{A^{2}(p) p^{2} \! + \! ({\sigma}^{2} + \vec{\pi}^{2}) B^{2}(p)}\right] 
+ \frac{ ({\sigma}^{2} + \vec{\pi}^{2} -1 ) B^{2}(p) } {A^{2}(p)
{p}^{2} \! + \! B^{2}(p)} \right\} \, .
\end{equation}
\end{widetext}
With the correction from the motion of center of mass, the
zero-point effect, and the color-electronic and color-magnetic
interactions being taken into account, the total energy of a bag
(involving three valence quarks) is given as
\begin{equation}
E_{B}(R) = 3\varepsilon_{j}(R)+\frac{4}{3}{\pi}
R^{3}{\mathscr{B}}-\frac{Z_0}{R} \, ,
\end{equation}
where $\varepsilon_{j}(R)$ is the energy eigenvalue of the quark's
equation of motion, $R$ is the radius of the bag, $ Z_{0}/R$ denotes
the contribution of the corrections of the motion of the center of
mass, zero-point energy, and other effects with $Z_{0}$ being a
parameter.
Just the same as that in Ref.~\cite{GCM1}, the bag is identified as
a nucleon in the present work. With the equilibrium condition
$$ \frac{d{E_{B}^{}}(R)}{dR} = 0 \, , $$
we can obtain the radius of a nucleon.

\section{Algorithm and Numerical Results }

\subsection{Algorithm}

From the description in last section, we know that the property of a
nucleon (for instance, its mass, radius and bag constant) is
determined by the solutions of the equations of motion of the quarks
and (chiral) mesons in the soliton. To solve the equations of
motion, one needs the solutions $A(p^{2})$ and $B(p^{2})$ of the
quark's Dyson-Schwinger equation. Then after solving the quark's DSE
in Eq.~(7), or more explicitly [with the help of the decomposition
in Eq.~(8)] the coupled equations
\begin{widetext}
\begin{eqnarray}
\begin{split}
(A(p^2)-1)p^2 & =  g^{2}C_F \int{\frac{d^4q}{(2\pi)^4}
D(p-q)tr\left[i\gamma\cdot{p}\ t_{\mu\nu}\gamma_\mu
S(q) \Gamma_\nu(p,q)\right]}  \, , \\
B(p^2)-m_0 & =  g^{2}C_F
\int{\frac{d^4q}{(2\pi)^4}D(p-q)tr\left[t_{\mu\nu}\gamma_\mu S(q)
\Gamma_\nu(p,q)\right]}  \, ,
\end{split}
\end{eqnarray}
\end{widetext}
where $C_{F}$ is the eigenvalue of the quadratic Casimir operator in
the fundamental representation of the color symmetry group [for
$SU(N_c)$, $C_{F}=(N_c^2-1)/2N_c$, it reads $4/3$ at $N_c=3$].

To investigate the temperature dependence of the property of a
nucleon, one should at first discuss the form of the quark's DSE at
finite temperature.

It has been well known that the appearance of (nonzero) temperature
$T$ in the QCD reduces the $O(4)$ symmetry to $O(3)$. Thus the
quark's four-momentum $p$ should be rewritten as $ p = (\vec{p},
{\omega}_{n})$, where ${\omega}_{n} = (2n+1) {\pi} T$ $(n\in Z)$ are
the discrete Matsubara frequencies of the quark, and the
four-dimensional integral needs to be replaced by~\cite{Kapusta}

\begin{equation}
\int{\frac{d^4p}{(2\pi)^4}} \to
T\sum_{n=-\infty}^{\infty} {\int{\frac{d^3p}{(2\pi)^3}}} \, .
\end{equation}
The decomposition of the dressed quark propagator needs to be
rewritten as
\begin{equation}
S^{-1}\!(\vec{p},{\omega_{n}^{}}\! ) = i\vec{\gamma}\! \cdot \!
\vec{p} A(|\vec{p}|,{\omega_{n}^{}}\! ) + i{\gamma_{4}^{}}
{\omega_{n}^{}} C(|\vec{p}|,{\omega_{n}^{}} \! )
 +  B(|\vec{p}|,{\omega_{n}} \! ) \, .    \;
\end{equation}
Furthermore, the gluon propagator at finite temperature can be
generally expressed as~\cite{Kapusta}
\begin{equation}
D_{\mu\nu}(\vec{k},\omega_n)=\frac{D_{T}(\vec{k},\omega_n)}{k^2}
P_{\mu\nu}^{T}(k)+\frac{D_{L}(\vec{k},\omega_n)}{k^2}P_{\mu\nu}^{L} (k) \, ,
\end{equation}
with the transverse and longitudinal projectors
\begin{eqnarray}
\begin{split}
P_{\mu\nu}^{T}(k) &= \left\{
\begin{array}{ll} \big(\delta_{ij}-\frac{k_ik_j}{\vec{k}^{2}}\big)
\delta_{i\mu}\delta_{j\nu}, & \;\; {\mu}, {\nu} =1,2,3, \\
0, & \;\; {\mu} \; {\textrm{or/and} } \; {\nu} = 4 ,  \\
\end{array} \right.
\\
P_{\mu\nu}^{L}(k) &= \left(\delta_{\mu\nu}-\frac{k_\mu
k_\nu}{{k}^{2}}\right)
-P_{\mu\nu}^{T}(k) \, . \\
\end{split}
\end{eqnarray}

Due to the lack of detailed information about the gluon propagator
at finite temperature, one usually na\"{i}vely assumes that the
transverse and longitudinal parts of the gluon propagator are equal
and independent of temperature (see, for example,
Ref.~\cite{Fischer2009}), i.e., one has approximately
$D_{T}(\vec{k},\omega_n)=D_{L}(\vec{k},\omega_n)=D(k) $. In this
assumption and bare vertex approximation, we get three coupled
integral equations for the functions $A(\vec p,\omega_n)$, $C(\vec
p,\omega_n)$ and $B(\vec p,\omega_n)$ as
\begin{widetext}
\begin{eqnarray}
\begin{split}
A(|\vec{\mathstrut p}|,\omega_n) & = 1+g^2T\frac{C_F}{\vec{p}^{\ 2}}
 \sum_m{\int{\frac{d^3\vec{q}}{(2\pi)^3}{\frac{D(k)}{k^2}}
 {\frac{1}{\Delta}}\left\{[(\vec{\mathstrut p}\cdot \vec{\mathstrut q})k^2
 +2(\vec{\mathstrut p}\cdot \vec{\mathstrut k})
 (\vec{\mathstrut q}\cdot \vec{\mathstrut k})]
 A(|\vec{\mathstrut q}|,\omega_m)+2{\omega_{m}} {\Omega_k}(\vec{\mathstrut p}
 \cdot \vec{\mathstrut k})C(|\vec{\mathstrut q}|,\omega_{m})   \right\}  }} \, , \\
C(|\vec p|,\omega_n)&=1+g^2T\frac{C_F}{\omega_n^2}
\sum_m{\int{\frac{d^3\vec{q}}{(2\pi)^3}{\frac{D(k)}{k^2}}{\frac{1}{\Delta}}
\left\{[\omega_n\omega_mk^2+2\omega_n\omega_m\Omega_k^2]
C(|\vec{\mathstrut q}|,\omega_m) + 2 {\omega_{n}}
{\Omega_k}(\vec{\mathstrut p}\cdot \vec{\mathstrut k})
A(|\vec{\mathstrut q}|,\omega_{m})   \right\}  }} \, , \\
B(|\vec p|,\omega_n)&=m_0+g^2TC_F
\sum_m{\int{\frac{d^3\vec{q}}{(2\pi)^3}
{\frac{D(k)}{\Delta}}3B(|\vec q|,\omega_{m})} }   \, ,
\end{split}
\end{eqnarray}
\end{widetext}
where $k^2\equiv\vec{k}^2+\Omega_k^2$, $\Omega_k
\equiv\omega_n-\omega_m$, and $\Delta=\vec q^{\ 2}A^2 +
\omega_{m}^{2} C^{2} + B^{2} $.

As mentioned in last section, in view of the non-topological soliton
ans\"{a}tz, one can take a nucleon as a soliton bag. In the chiral
limit ($m_{0} =0$), one has found that there exist two types of
solutions for the quark's DSE. One is the Nanmbu-Goldstone solution
which corresponds to the chiral symmetry spontaneously broken phase.
The other is the Wigner solution which represents the state with the
chiral symmetry. One gets then the effective bag constant as the
pressure difference between the Nambu-Goldstone solution and the
Wigner solution, which reads
\begin{widetext}
\begin{eqnarray}
\begin{split}
\mathscr{B}(T)&\equiv P[\mathcal {G}^{NG}]-P[\mathcal {G}^{W}]\\
&=4N_c \sum_{m}^{}\int{\frac{d^3p}{(2\pi)^3}}
\left\{ln\left[\frac{\Delta_{NG}}{\Delta_{W}}\right]
+\frac{\vec{p}^{2}A_{NG} + {\omega}^{2}_{m} C_{NG} }{\Delta_{NG}}
-\frac{\vec{p}^{2}A_{W} + {\omega}^{2}_{m} C_{W} }{\Delta_{W}}
\right\} \, ,
\end{split}
\end{eqnarray}
\end{widetext}
where $\Delta_{NG} \equiv \vec{p}^{2} A_{NG}^{2} + {\omega}_{m}^{2}
C_{NG}^{2} + B_{NG}^{2}$, $\Delta_{W} \equiv \vec{p}^{2} A_{W}^{2} +
{\omega}_{m}^{2} C_{W}^{2}$, and $A_{NG}$, $C_{NG}$, $B_{NG}$,
$A_{W}$, $C_{W}$ denotes the Nambu-Goldstone solution, the Wigner
solution for the DSE, respectively.

With the solutions of the DSE as input, we can have the explicit
expression of the quark's equation of motion at finite temperature
as
\begin{widetext}
\begin{equation}
[i\vec{\gamma}\cdot {\vec{p}}A(\vec{p},T) + i {\gamma}_{4}
{{\omega}_{n}} C(\vec{p},T) + B(\vec{p},T) ] u_{j}(\vec{p},T) +
\int\!\!{\frac{d^3k}{(2\pi)^3}}B\left(\frac{p\!+\! k}{2},T\right) 
\left[\hat\sigma(\vec{p}-\vec{k})
+i\gamma_5\vec{\tau}\cdot\vec{\pi}(\vec{p}-\vec{k})\right]u_j(\vec{k},T)=0
\, ,
\end{equation}
\end{widetext}
where $\hat\sigma\equiv\sigma - 1$. After solving the related
eigenequation we can obtain the eigenenergy of the quark at finite
temperature, $\varepsilon_{j}(T)$.
One can in turn study the properties of a nucleon at finite
temperature by extending the GCM soliton model described in last
section.

In the case of zero temperature, one usually takes only the lowest
energy for the $\varepsilon_{j}(R)$ in Eq.~(19). At
finite-temperature, due to the influence of temperature, the
contribution of the excited states of the quarks must be included.
Then the energy of the bag, i.e., the mass of a nucleon, at finite
temperature $T$ should be written as
\begin{equation}
M(T) = E_{B}(T) = 3\overline{\varepsilon_{j}(T)} - \frac{Z_0}{R}
+\frac{4}{3}\pi R^3\mathscr{B}(T) \, ,
\end{equation}
where $\overline{\varepsilon_{j}(T)}$ is the average of a quark's
energies at all possible states.

In the spirit of the most simple approximation of the
GCM~\cite{GCM1} (mentioned at the end of the last section), the
quarks in the bag (GCM soliton) at finite temperature can also be
regarded as the free one which satisfies the Dirac equation, and the
energy eigenvalue can be expressed as
\begin{equation}
\varepsilon_{j}(T)=\frac{\kappa_{j}}{R(T)} \, ,
\end{equation}
where $j$ denotes the quantum number labeling the energy level.
Since quarks are fermions, we take the Fermi-Dirac statistics to
evaluate the average energy of each quark in the bag at finite
temperature, and have
\begin{eqnarray}
\overline{\varepsilon_{j}(T)} = N
\sum_{j=0}^{\infty}{\frac{\varepsilon_{j}(T)}{1+e^{\varepsilon_{j}(T)/T}}}
\, ,
\end{eqnarray}
where $N$ is the degeneracy of quarks.

Then, by solving the stability condition $\frac{d M(T)}{dR} = 0 $,
i.e.,
\begin{eqnarray}
\frac{dE_B(T)}{dR}=3\frac{d{\overline{\varepsilon_{j}(T)}}}{dR} +
\frac{Z_{0}}{R^2} + 4\pi R^2\mathscr{B}(T) = 0 \, ,
\end{eqnarray}
we can obtain the (stable) nucleon radius $R(T)$ and the mass of a
nucleon (the energy of the bag) $M(T)$.

It is apparent that after solving the quark's DSE and in turn the
equations of motion of the quark and meson fields at zero and
nonzero temperature, we can obtain the bag constant, the mass and
the radius of a nucleon in the corresponding circumstance and
discuss the variation characteristic of the property with respect to
temperature. To solve the quark's DSE, we take a simplified form of
the effective gluon propagator in Ref.~\cite{Maris1997}
\begin{equation}
g^2D(k)=4{\pi}^{2} D \frac{k^{2}}{{\omega}^{6}} \exp \left(
-\frac{k^{2}}{{\omega}^{2}} \right) \, ,
\end{equation}
where $D$ and $\omega$ are dimensional parameters that can be
determined by fitting empirical date. Such an effective gluon
propagator is naturally an extension of the Munczek-Nemirovsky
model~\cite{MNmodel} and consistent with those given in lattice QCD
calculations (see, for instance,
Refs.~\cite{Sternbeck2005,Bowman2007}) and solving the coupled DSEs
of quark, gluon, and ghost (see, for instance,
Ref.~\cite{Alkofer2004}). It has also been shown to be successful in
describing many hadron
properties~\cite{DSE-Roberts,Maris1997,Alkofer2002,Roberts2008,Lei2009}.

\subsection{Calculation and Numerical Results}

We first solve the quark's DSE at chiral limit ($m_{0}^{} = 0 $) and
zero temperature with parameters $D = 1.0 \, \textrm{GeV}^{2}$ and
$\omega = 0.5 \, \textrm{GeV}$, which have been widely used (see,
for example, Refs.~\cite{Roberts2008,Lei2009}). The obtained result
of the Nambu-Goldstone solution of the DSE at zero temperature is
displayed in Fig.~1.
It shows evidently that our result reproduces exactly that given in
Ref.~\cite{Alkofer2002}.
\begin{figure}[htb]
\centering
\includegraphics[width=8.0cm]{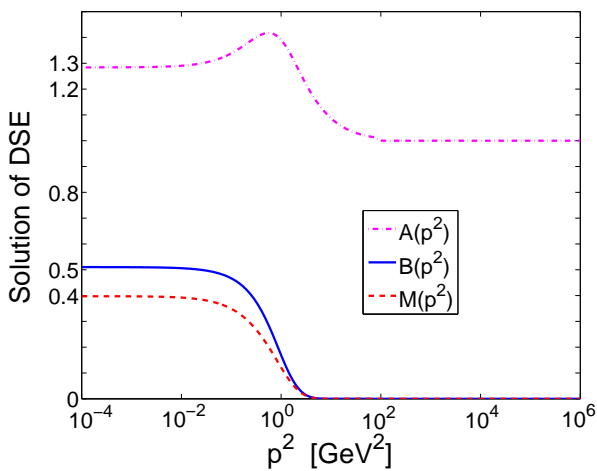}
\caption{(Color online) Calculated result of the Nambu-Goldstone
solution of the quark's DSE at zero temperature, with parameters in
the effective gluon propagator $\omega=0.5\, \textrm{GeV}$, $D = 1\,
\textrm{GeV}^{2}$.}
\end{figure}

\begin{figure}[htb]
\centering
\includegraphics[width=8.0cm]{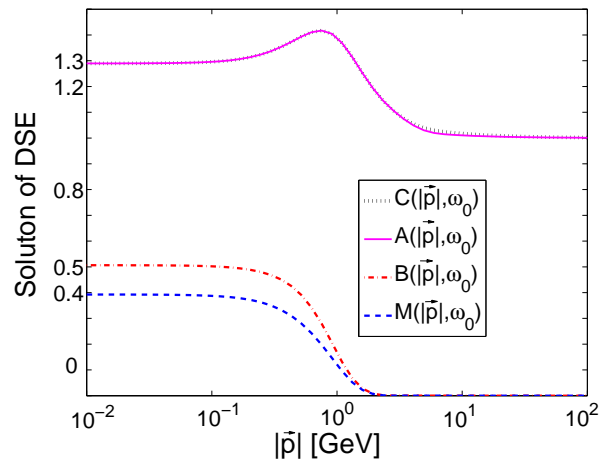}
\caption{(Color online) Calculated results of the functions
$A(|\vec{p}|,T)$, $C(|\vec{p}|,T)$, $B(|\vec{p}|,T)$ and
$M(|\vec{p}|,T)$ in the Nambu-Goldstone solution of the quark
propagator at a temperature $T=30$~MeV, with parameters in the
effective gluon propagator $\omega=0.5\, \textrm{GeV}$, $D = 1 \,
\textrm{GeV}^{2}$.}
\end{figure}

We then solve the quark's DSE at nonzero temperature with the same
effective gluon propagator.
Figure~2 illustrates the obtained results of the functions
$A(|\vec{p}|,T)$, $C(|\vec{p}|,T)$, $B(|\vec{p}|,T)$ in the
Nambu-Goldstone solution and the mass function $M(|\vec{p}|,T)$ at a
temperature $T=30$~MeV, as an example of those at nonzero
temperature.
We find from the figure that, just as expected, as the temperature
is lower, the functions $A(|\vec{p}|,T)$, $C(|\vec{p}|,T)$,
$B(|\vec{p}|,T)$ and $M(|\vec{p}|,T)$ have the correct zero
temperature limit. Hence, even though we have not included
explicitly the temperature effect in the effective gluon propagator,
the calculated result can demonstrate the temperature dependence of
quark propagator with quite high precision. In addition, it can be
noticed from Fig.~2 that functions $A(|\vec{p}|,\omega_0)$ and
$C(|\vec{p}|,\omega_0)$ have the same behavior when the temperature
is low, just as that of the quark propagator at low chemical
potential~\cite{Chen2008}.
To study the Matsubara frequency dependence of the quark propagator,
we illustrate the variation behaviors of the functions
$A(|\vec{p}|,\omega_{n})$ and $B(|\vec{p}|,\omega_{n})$ at a
temperature $T=30$~MeV in Fig.~3 [since Fig.~2 has shown that the
functions $A(|\vec{p}|,\omega_n)$ and $C(|\vec{p}|,\omega_n)$ at
$T=30$~MeV are almost exactly equal to each other, we do not show
the function $C(|\vec{p}|,\omega_n)$] as a representative.
From Fig.~3, we can find that, when the temperature is low, both
functions $A$ and $B$ involve an approximate bilateral symmetry
about $n$ [since for a fixed temperature, $\omega_{n}$ is
proportional to $n$ due to the definition $\omega_{n} = (2n +1)
{\pi} T$], and the function $B$ decreases rapidly as the Matsubara
frequency increases. It gives us a posterior knowledge that when we
carry out the summation of the Matsubara frequencies $\omega_{n}$,
it is not necessary to do that up to a very large number of $n$.

\begin{figure}[htb]
\centering
\includegraphics[width=8.0cm]{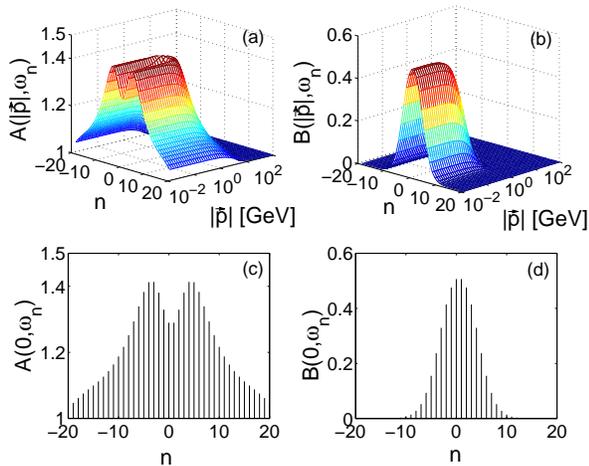}
\caption{(Color online) Calculated results of the functions
$A(|\vec{p}|,\omega_{n})$ and $B(|\vec{p}|,\omega_{n})$, where
$\omega_{n}=(2n+1)\pi T$, in the Nambu-Goldstone solution of the
quark propagator in chiral limit and at a temperature $T=30$~MeV
[(a), (b), respectively] and the special case at $| \vec{p} | =0 $
[(c), (d), respectively]. The calculations are also carried out with
parameters $\omega=0.5\, \textrm{GeV}$, $D = 1 \, \textrm{GeV}^{2}$
in the effective gluon propagator. }
\end{figure}

\begin{figure}[htb]
\centering
\includegraphics[width=8.0cm]{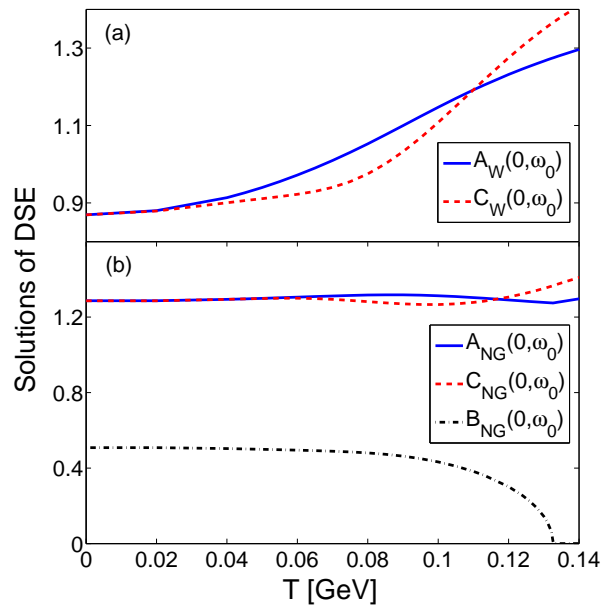}
\caption{(Color online) Calculated results of the temperature
dependence of the Wigner solution of the quark's DSE at zero
momentum and zero mode Matsubara frequency [i.e., the functions
$A_{W}(|\vec{p}|=0,{\omega_{0}^{}})$ and
$C_{W}(|\vec{p}|=0,{\omega_{0}^{}})$] (panel (a)) and the
Nambu-Goldstone solution under the same conditions [i.e., the
functions $A_{NG}(|\vec{p}|=0,{\omega_{0}^{}})$,
$C_{NG}(|\vec{p}|=0,{\omega_{0}^{}})$ and
$B_{NG}(|\vec{p}|=0,{\omega_{0}^{}})$] (panel (b)). The calculations
are also carried out with parameters $\omega=0.5\, \textrm{GeV}$, $D
= 1 \, \textrm{GeV}^{2}$ in the effective gluon propagator. }
\end{figure}

To demonstrate the temperature dependence of the quark propagators
explicitly, we illustrate the calculated variation behaviors of the
Wigner solution at zero momentum and zero mode Matsubara frequency
[i.e., the functions $A_{W}(|\vec{p}|\! = \! 0,\omega_{0})$ and
$C_{W}(|\vec{p}|\! = \! 0,\omega_{0})$] and the Nambu-Goldstone
solution under the same conditions [i.e., the functions
$A_{NG}(|\vec{p}|\! = \! 0,\omega_{0})$, $C_{NG}(|\vec{p}| \! = \!
0,\omega_{0})$ and $B_{NG}(|\vec{p}| \! = \! 0, \omega_{0})$] with
respect to temperature in Fig.~4 as a representative.
From Fig.~4, we can find that, when the temperature is low, the
functions $A_{NG}$ and $C_{NG}$ coincide with each other very well.
So do the $A_{W}$ and $C_{W}$ except that the temperature for the
deviation between them to appear is lower. It indicates that only as
the temperature is quite high, the breaking from O(4) symmetry to
O(3) symmetry (especially, for the physical state, i.e., the
Nambu-Goldstone state) becomes obvious. Moreover, the decreasing
feature of the function $B_{NG}$ is a manifestation of gradual
restoration of chiral symmetry.

\begin{figure}[htb]
\centering
\includegraphics[width=8.0cm]{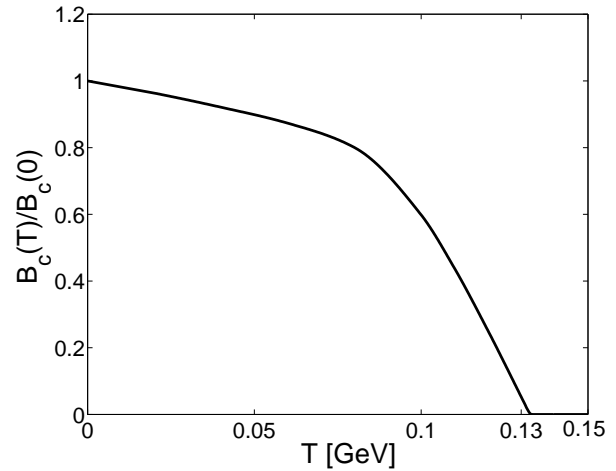}
\caption{Calculated result of the variation behavior of the bag
constant of a nucleon with respect to temperature.}
\end{figure}

With the solutions of the quark's DSE as input, we solve the
equations of motion of the quark and the chiral meson fields, and
then obtain the property of a nucleon at zero and nonzero
temperature. The obtained property of a nucleon at zero temperature
is bag constant $\mathscr{B}(0) = (162\, \textrm{MeV})^{4}$, mass
$M(0)=939$~MeV (with the parameter $Z_{0}$ is fixed as 3.08), and
radius $R(0)=0.85$~fm.
It should be noted that the presently fixed value of the parameter
$Z_{0}$, $3.08$, is larger than the usually taken one, $1.84$. Such
a large value arises from the fact that the term $-{Z_{0}}/R$ is a
combination of the contributions from not only the zero-point energy
but also those of the color-electronic and color-magnetic
interactions, the correction on the motion of center-of-mass and
other effects. And it is consistent with the most recent
result~\cite{Caillon2010} and our previous results~\cite{LYX200135}.
The gained variation behaviors of the nucleon's bag constant, mass,
and radius with respect to temperature are illustrated in Figs.~5,
6, 7, respectively. From Figs.~5--7, one can notice that, with the
increasing of temperature if it is below a critical one, the bag
constant and the mass of a nucleon decrease, and the radius of the
nucleon increases. At critical temperature $T=133$~MeV, the bag
constant and the mass of the nucleon vanish and the radius tends to
be infinite. It manifests that the nucleon can no longer exist as a
bag soliton, so that the quark deconfinement happens.
Admittedly, such a obtained critical temperature may be
model-dependent, however, the gradual variation features of the
nucleon's property indicate that the deconfinement phase transition
process is that, with the increase of temperature, the nucleons in
the matter touches with each other at first due to the increase of
the radius, then the fields of the ingredients of the nucleons mixed
with each other and the bound strength gets weaker simultaneously.
As the bound (the bag constant) vanishes, the deconfinement phase
transition occurs. Therefore, the temperature driven deconfinement
process may be, in fact, a crossover but not a low order phase
transition.

\begin{figure}[htb]
\centering
\includegraphics[width=8.0cm]{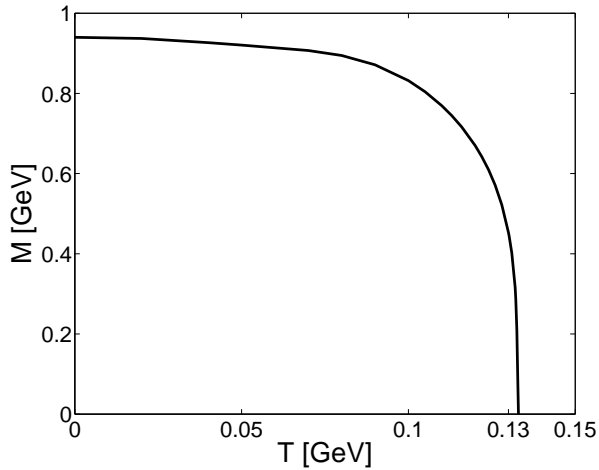}
\caption{Calculated result of the variation behavior of the mass of
a nucleon with respect to temperature.}
\end{figure}

\begin{figure}[htb]
\centering
\includegraphics[width=8.0cm]{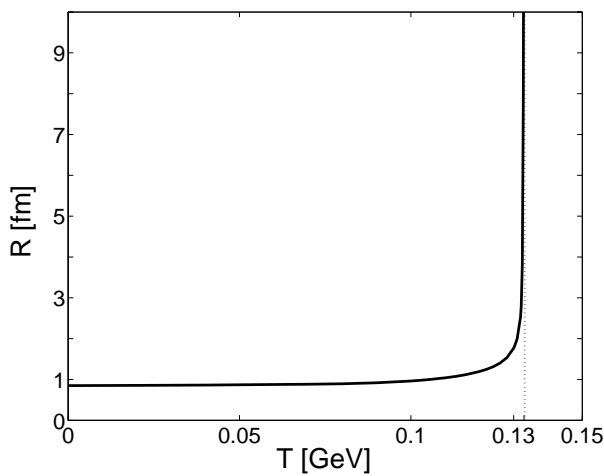}
\caption{Calculated result of the variation behavior of the radius
of a nucleon with respect to temperature.}
\end{figure}

When discussing the QCD phase transition, one usually interests in
the chiral symmetry and its dynamical breaking, too, and takes the
chiral quark condensate as a order parameter in the case of chiral
limit, which is defined as
\begin{equation}
-\langle \bar{q} q\rangle=N_c T\sum_{n=-\infty}^{n=+\infty}
\int\frac{d^3p}{(2\pi)^3}tr[S(\vec{p},\omega_n)] \, .
\end{equation}

We then calculate the temperature dependence of the chiral quark
condensate.
The obtained result is shown in Fig.~8. The figure displays
evidently that the chiral quark condensate decreases gradually with
the increase of temperature and vanishes at a critical temperature.
It indicates that dynamical chiral symmetry breaking effect gets
weaker and weaker with the increase of temperature and the chiral
symmetry can be restored as the temperature reaches the critical
one.
In the case with parameters $\omega=0.5\, \textrm{GeV}$, $D = 1 \,
\textrm{GeV}^{2}$ in the effective gluon propagator, the critical
temperature for the chiral symmetry to be restored is also
approximately $133$~MeV, the same as that for the deconfinement.

\begin{figure}[htb]
\centering
\includegraphics[width=8.0cm]{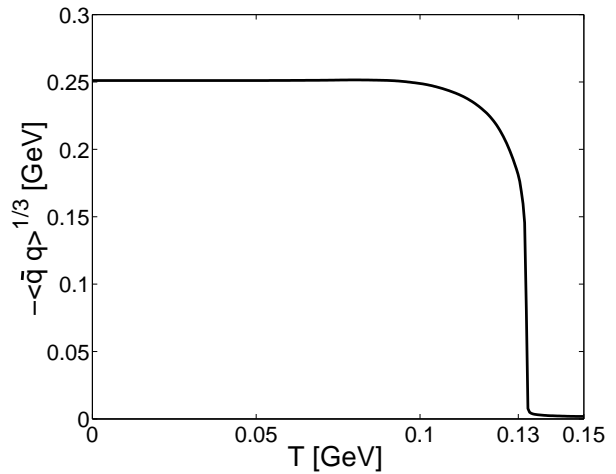}
\caption{Calculated result of the variation behavior of the chiral
quark condensate with respect to temperature.}
\end{figure}

\begin{figure}[htb]
\centering
\includegraphics[width=8.0cm]{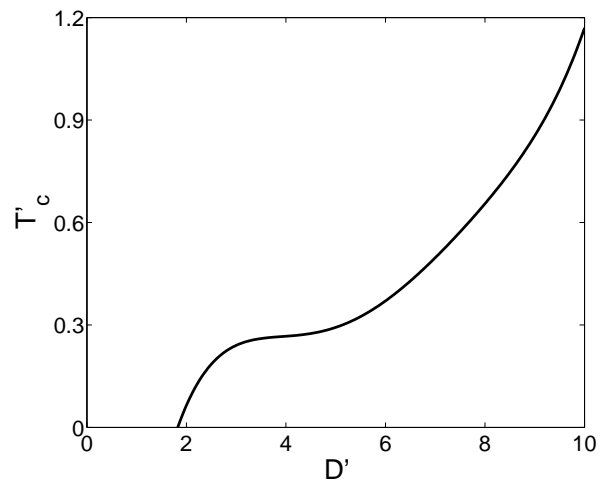}
\caption{Calculated result of the variation behavior of the scaled
critical temperature with respect to the scaled strength parameter
$D'$ in the effective gluon propagator, where $T'_{c} =
T_{c}/\omega$ and $D'=D/\omega^2$ are the scaled quantities.}
\end{figure}

As mentioned above, the critical temperature for the QCD phase
transitions to happen may be parameter dependent. To demonstrate the
parameter dependence explicitly, we carry out a series calculations
with various values of the coupling strength parameter $D$ and the
screening width parameter $\omega$ in the effective gluon
propagator. Due to the good behavior of the Gaussian-type gluon
propagator, we can scale the $T_c$ and $D$ by $\omega$ with
definition $T'_{c} \equiv T_{c}/\omega$ and $D' \equiv
D/\omega^{2}$. The obtained variation behavior of the $T'_{c}$ with
respect to the $D'$ is displayed in Fig.~9. One can find easily from
the figure that the critical temperature increases when the coupling
strength gets larger.
Furthermore, there exists a critical scaled coupling strength $D'$,
below which the critical temperature for the deconfinement maintains
zero. In fact, below the critical coupling strength, the quark's DS
equation does not have a Nambu-Goldstone
solution~\cite{YuanWei,ChangLei}. In other words, there exists only
quarks in chiral symmetry. In turn, we have only deconfined quarks,
but no nucleons (more general, hadrons) even if the temperature is
zero.

\section{Summary and Remarks}

In this paper we have calculated the temperature dependence of the
quark propagator by solving the quark DSE with a Gaussian-type
effective gluon propagator. Based on the calculations, we
investigated the temperature dependence of the bag constant, the
mass and the radius of a nucleon in the framework of the GCM soliton
model. It shows that, as the temperature is lower than a critical
value, the bag constant and the mass decrease and the radius
increases with the increasing of the temperature. In the case with
parameters $\omega=0.5\, \textrm{GeV}$, $D = 1 \, \textrm{GeV}^{2}$
in the effective gluon propagator, the critical temperature is found
to be about 133~MeV. At the critical temperature, the bag constant
and the mass decrease to zero and the radius increases to infinity.
It means that the nucleon can no longer exist as a bag soliton, so
that the deconfinement phase happens. This indicates evidently that
the quark deconfinement phase transition can take place at high
temperature. Moreover, we give the dependence of the critical
temperature on the interaction strength parameter $D$ in the
effective gluon propagator. It shows that, as the interaction
strength parameter is larger than a critical value, the critical
temperature increases with the increasing of the strength parameter.

Even though the temperature dependence of some of the properties of
nucleon is given with some approximations and model parameters in
the present work, the qualitative behavior would be universal and it
is the first one given with quite a sophisticated approximation of
QCD. Of course, there are various aspects to be improved. For
example, we take the commonly used effective gluon propagator, which
is independent of temperature, to solve the quark DSE and make use
of the preliminary GCM soliton model~\cite{GCM1,GCM2,LYX200135}. It
is necessary to implement the real GCM soliton
model~\cite{GCMsoliton1,GCMsoliton2} with solving at first the
coupled DSEs of the quark, gluon and ghost, and then the coupled
equations of the quark and the chiral fields, with the inclusion of
the temperature effect. In more detail, for the quark gluon
interaction vertex, we take simply the bare vertex $\gamma_{\mu}$ in
our present work, in fact more realistic vertex functions, such as
the BC vertex~\cite{BC-vertex}, which has been shown to be able to
improve the calculation of meson properties greatly~\cite{Lei2009},
even the BC vertex together with the transverse part being included
simultaneously~\cite{CP-vertex,HeHX,CRprivate}, should be
implemented to make the calculation with much more solid QCD
foundation.
It is also necessary to notice the dimensionless quantity $Z_{0}$
which takes account of the contributions of the zero-point effect,
the color-electronic and color-magnetic interactions, the motion of
center-of-mass and all the others. In our present work, we handle
it, in the commonly taken way, as a free parameter to be fixed by
the property of a nucleon in free space. In fact, all the aspects of
the $Z_{0}$ are quite complicated and have recently been paid great
attentions (see for example, Refs.~\cite{Gutsche2000,Oxman20057}
tried to evaluate the zero-point effect part from the gluon field
fluctuations directly). Furthermore, extending the result of the
thermal Casimir effect in ideal metal rectangular
boxes~\cite{Geyer2008}, we infer that the $Z_{0}$ (at least, the
zero-point effect part) may depend on temperature. It would then be
interesting to study the temperature dependence of the parameter
$Z_{0}$.
The related investigations are in progress.

\section*{Acknowledgements}

This work was supported by the National Natural Science Foundation
of China under contract Nos. 10425521 and 10935001, the Major State
Basic Research Development Program under contract No. G2007CB815000.
Helpful discussions with Dr. Lei Chang are acknowledged with great
thanks.



\end{document}